\documentclass[runningheads,a4paper]{llncs}

\usepackage{amssymb}
\usepackage{graphicx}
\usepackage{float}
\usepackage{url}

\begin{document}

\mainmatter  

\title{Microservices as an Evolutionary Architecture of
Component-Based Development: \\A Think-aloud Study}

\titlerunning{Running title}

%
%
\author{Reza M. Parizi%
\thanks{Corresponding author. Tel: +1-470-578-2118;
Fax: +1-470-578-9219. 
}%
\
}

\authorrunning{}

\institute{{Department of Software Engineering and Game Development\\ Kennesaw State University, Marietta, GA 30060, USA\\ {\fontfamily{qcr}\selectfont rparizi1@kennesaw.edu}\\}}

%
%

\toctitle{Lecture Notes in Computer Science}
\tocauthor{Authors' Instructions}
\maketitle

\begin{abstract}
Microservices become a fast growing and popular architectural style based on service-oriented development. One of the major advantages using component-based approaches is to support reuse. In this paper, we present a study of microservices and how these systems are related to the traditional abstract models of component-based systems. This research focuses on the core properties of microservices including their scalability, availability and resilience, consistency, coupling and cohesion, and data storage capability, while highlighting their limitations and challenges in relation to components. To support our study, we investigated the existing literature and provided potential directions and interesting points in this growing field of research. As a result, using microservices as components is promising and would be a good mechanism for building applications that were used to be built with component-based approaches. 
\end{abstract}

\section{Introduction}

Microservices are the new rising architecture in the enterprise market, and mainly adopted by massively deployed applications such as Netflix, eBay, Amazon and even Uber \cite{1}. Microservice has similarity with component-based development (CBD) on objectives which is to create a large complex architecture from existing small reusable elements by assembling and replacing interoperable parts \cite{2}. The CBD (or often CBSD) idea was first introduced by Douglas McIlroy \cite{3} and it was intended to use components in the commercial production. Since after, many researches on CBD were conducted within last 20 years \cite{4}, \cite{5}. The definition of microservice goes an architectural pattern for development of distributed applications that builds an application as a collection of loosely coupled services or smaller independent components that implement business capability \cite{6}. Microservices communicate with lightweight mechanisms, often an HTTP resource API. As the demand for complex applications are increasing, driven with frequent changes in technology stack, developmental practices have evolved from the traditional "monolithic" architecture, which is all developed all in one piece, to a more fine-grained distributed architecture, most recently advocated by microservices. Benefits of microservices over monolithic architecture are scalability and reusability, as well as efficiency for the most part. 

Over the past two decades, industry needs have driven software architecture in various directions from client-server, service-oriented architecture (SOA) to microservices. As the technologies (e.g., CORBA, OSGi, Sun JavaBeans or Enterprise JavaBeans components) and traditional practices of CBSD are getting less popular and demanding, it makes more sense to study whether microservices could be used as components to surrogate these practices as promised. Inspired by this, the goal of this research paper is to explain microservices in the field of component-based software development (CBSD), analyzing their related properties to provide an overview of the topic and to lay a first-step basis for future research in the CBSD.

The rest of the paper is organized as follows. Section 2 gives an overview the related work. Section 3 contains the research design and synthesis of results. Sections 4 discusses briefly the main observations with respect to research questions, and outlines future directions. Finally, Section 5 summarizes concluding remarks.

\section{Related Work}

In this part, we overview the most current and existing researches around microservices architectures whose aims have been directed to assessing and evaluating works in the field.

Francesco et al. conducted a meta-study research on architecting microservices \cite{1}. Their research focuses on (i) the publication trends of research studies about architecting microservices (ii) the focus of research on architecting microservices, and (iii) the potential for industrial adoption of existing research on architecting microservices. They searched over 300 papers and selected 71 primarily studies. They concluded architecture analysis emerges as the most popular architecting activity and research scope, but supports for proper architecting of microservices are still immature.

Alshuqayran et al. performed a systematic mapping study on microservice architecture \cite{7}. They selected 33 articles published in between 2014 to 2016. Their study focuses on (i) the architectural challenges faced by microservice-based systems, (ii) the architectural diagrams used for representing them, and (iii) the involved quality requirements. Their research is somewhat similar in the techniques we used, however their research questions were different than ours.

Shadija et al. carried out a research on understanding of microservices \cite{6}. In their research, they approached by comparing microservices architecture with service-oriented architecture that identified key characteristics that will assist developers to choose the most suitable approach. They concluded that SOAPService is a communication layer on top of the business logic of an application while a microservice is a service-oriented architectural style for the application.

Sampaio et al. conducted a research on supporting microservice evolution \cite{8}. In this research, they outlined checking for upgrade consistency, identifying architectural improvements and evaluating changing deployment trade-offs. In their article, they concluded that microservices can perform a flexible and scalable approach of distributed business logic.

Most recently, Cerny et al. \cite{9} proposed a comprehensive survey-nature study contextualizing the main concepts behind both SOA and microservices. In this work, the authors explained the key differences between these two architectures and their features, and presented research and industry viewpoints and challenges on adoption of microservices compared to SOA.

In summary, our study of literature (including the aforementioned works) reveals that the concept of CBSD has not been discussed in the current literature related to microservices, which is the main center of discussion in this research.

\section{Research Design and Results}

The main objective of this study is to explain microservices in the field of component-based software development (CBSD), analyzing their related properties to provide an overview of the topic and to lay a first-step basis for future research in the CBSD. As a result, the main research question addressing this work is: \textit{can microservices be used as components in massively deployed and complex applications?} Since microservices can intuitively be used broadly for terms, we specify certain key features of microservices and discuss their relevance.

\subsection{Research Questions}
To support the main research question mentioned above, we formulated the following research questions to drive the study:\\

RQ1. What are the main motivations to study microservices in CBSD?

RQ2. What are the limitations of microservices in this field?

RQ3. What should be the future research in this field?

\subsection{Search Strategy and Selection}
The search process was conducted via electronic searching on online scientific databases. Therefore, first, we use the following electronic databases to find research papers that releate to "microservices" \& "architecture" \& "components" \& "component-based development" \& "component-based software development" \& "CBSD" in the followings:\\

- Elsevier (http://www.sciencedirect.com/)

- Springer (http://link.springer.com/)

- IEEE explorer (http://ieeexplore.ieee.org/)

- ACM digital library (http://dl.acm.org/)

- John Wiley (http://onlinelibrary.wiley.com)

- Sage (http://online.sagepub.com/)

- Taylor and forensic (http://www.tandfonline.com/)

- Emerald (http://www.emeraldinsight.com/)

- Inderscience (www.inderscience.com/)

- Word scientific (www.worldscientific.com/)

- Google Scholar (https://scholar.google.com/)\\

We started our research by viewing 127 published articles using the above search strings and digital libraries.We then reviewed 29 published article that matches our title, and we further narrowed down to 17 papers with related summaries. Of those, most of the articles were different in the way we structured our research questions. Hence, we refined the selection and used 15 full text articles. These articles were thoroughly reviewed and contrasted against the main objective in this paper. Finally, we selected 12 researches which were most closely related to our study.

\subsection{Defining Distributed Micro-Deployments (DμD)}

Delving into Micro-Deployments, we assume that the term can roughly equate to broad-ranging updates to small services, in that minor updates done on a near-constant rate of development are distributed among a broad range of end-users.  Furthering this definition, we can elaborate on the topic by looking at an example of such, in this case, the app Uber, developed by Uber Technologies Inc. and their deployment of micro-updates to their app and servers. They "developed Micro Deploy (known as μDeploy for short), as in-house deployment system that builds, upgrades, and rolls back services at Uber" \cite{10}.

\subsection{Data Extraction and Synthesis Results}
In this section, we examine the core properties of microservices, and discuss how they could be related to CBSD with respect to the selected studies.  

\subsubsection{Microservices for Expansive Scale.}
When we look at an expansive scale regarding DμD, what are we looking for? We assume that we want to see a micro-deployment reach a broad range of deployment hits, but still maintain a balance of usage.  In \cite{8}, we can see that "important feature of microservices is their ability to scale in/out by removing or creating microservices replicas as necessary. This causes the microservices instances to have a short lifetime, inducing further dynamism and complexity". They also go into detail on an example wherein they split a microservice into two components so that their underutilized services of the duo do not get replicated needlessly. We can then determine that an expansive scale is relevant to hosting and removing instances of the microservices as needed.

\subsubsection{Microservices for Expansive Domain.}
An expansive domain is ever important for a microservice due to the desire for microservice providers to profit from hosting it. There are pros and cons to this: a larger domain means more income, more hits, but also means more traffic and more end-user issues. In \cite{6}, they remark upon the microservice's end goal: business capability. They further add that such domains like the medical or Internet infrastructure are domains to always consider.

\subsubsection{Microservices for Expansive Release Schedules.}
Furthering the research, an expansive release schedule is also desired. More updates mean more functionality, less issues, and more knowledge of what users want and approve of in your updates. However, this also poses issues in development, as more developers are pushed to make deadlines rushed, releases frequent, and can impose other issues on end-users like reaching data-caps or possible server outages.

Continuous development starts first with continuous integration that lets developers integrate their work with others work early and regularly. Typically, microservices are deployed using virtualization which is very cost ineffective and creates overhead \cite{13}. A possible solution to this by using containers that isolates and manages overhead much better than virtualization

Another method that has been researched is using cloud microservice to that of deploying a monolithic web application. Deployment for the microservice typically requires configuration with the application and services. New versions of microservices could easily break dependent services which made service versioning very important. Coordination between microservice teams is vital to preventing problem from occurring. The infrastructure cost for the microservice deployment compared to monolithic deployment is cheaper than the monolithic application but still has its own problems as previously mentioned. 

\textit{In traditional CBSD, instantiation and parallelism of several instances of one component is not literally supported, or it is challenging due to limited support from most languages. Thus, the scalability, domain-expansive friendly, and shorter release features of microservices can enhance the way the components were used in CBSD (a more distributed CBSD), and more legit reasons to use microservices as components.}

\subsubsection{The CAP Theorem.}

The CAP Theorem was first described by Eric Brewer in 1998. The basis for the CAP Theorem is that a distributed system can only guarantee two of the following three concerns: Consistency, Availability, and Partition tolerance \cite{11}.

\textit{Consistency:}
Gilbert and Lynch \cite{11} describe consistency as "the property that each server returns the right response to each request." By that definition, a system is consistent when all of the nodes can always agree on the same set of data across each partition. To guarantee consistency, one must sacrifice availability. This sacrifice exists because consistency requires an invested party to wait to give other invested parties the opportunity to finish posting answers if they are already doing so. This generally manifests itself as locking, which works well on a local machine. The more distributed a system, the more costly the communication; this means that locking has a greater impact on system performance and is thus less desirable.

\textit{Availability:}
Gilbert and Lynch describe availability as "the property that each request eventually receives a response" \cite{11}. It is often important to ensure a service can always provide a response, and it may be more and more important concern than consistency. Often this manifest itself in the form of message queueing and possibly even event-sourcing. The focus of systems that prioritize availability is often to always be able to eventually get the correct answer, but to never turn down clients or make them wait when the data is still in flux.

\textit{Partition Tolerance:}
Gilbert and Lynch describe partition tolerance as "communication among the servers is unreliable, and the servers can be partitioned into multiple groups that cannot communicate with one another" \cite{11}. This comes down to a form of reliability for the data in the sense that all the data must eventually converge to a point of consistency. In distributed systems, this is considered a requirement. It is therefore appropriate to conclude that partition tolerance is always a given, leaving an existing tradeoff between consistency and availability. This tradeoff is often decided on a per-problem basis and must be carefully considered at each design decision.

\textit{Microservices basically provides more availability than components in CBSD, while in terms of consistency, components could deliver more benefits over microservices due to each component provides a firm module boundary. While there could be a trade-off between availability and consistency, using microservies as components in more complex and user-centric applications would be still more prefered. The reason is the importance of resiliency and high availability of microservices in modern software applications.}

\subsubsection{Cohesion and Coupling.}
One of the major advantages of microservices is their tendency to create system with more cohesion and a lower degree of coupling. Heffner argues microservices are inherently cohesive and uncoupled because systems that do not fit this description are not microservices. He says that "a unit's size fits with the microservices concept if it is highly cohesive, with loose coupling to other units" \cite{12}.

\textit{Cohesion:} According to Heffner, "high cohesion is achieved by keeping together in one unit the things that change together" \cite{12}. A microservice must be a cohesive unit for it to be reasonably managed. If changes frequently require most services in the system to be changed, then a microservices pattern will only slow down production.

\textit{Coupling:} Hefner describes how "low coupling is achieved by designing so as to maximize the degree to which unit A's implementation can change without affecting the design or implementation of other units that refer to unit A" \cite{12}. This is especially useful for a microservice because of its need to be "hot-swappable": if a service is to be replaced with a new one, it is important other services working around this service are not inherently dependent only on that service to function correctly.

\textit{Microservices are less coupled than components. As pointed out by Fowler \cite{13}, using services as components would lead to a more explicit remote call mechanism. The reason is that most supporting languages in CBSD do not have a good strategy for defining an explicit Published Interface. Frequently, it's only documentation and discipline that prevents clients breaking a component's encapsulation, leading to overly-tight coupling between components.}

\subsubsection{ Data Storing Strategies}
Having each service be its own component allows developers more freedom in choosing ways to store their data. In this part, we discuss the various data storing applications and when to use each one in microservices domains.
There are many various databases available today. The main categories now are SQL, NoSQL, and most recently NewSQL.

Using a traditional SQL database will allow to organize data neatly into tables. The advantage of this method is that everything will be neatly structured, but it takes time to develop. A recent issue with traditional table SQL databases is with scalability. 

A NoSQL database is different from the traditional SQL databases. It was specifically made with the idea of having multiple devices allowing access to the data. One of NoSQL's advantages is that is non-relational, and this means that there are no tables.  Most NoSQL databases are open source. One disadvantage however is that there is no standardization. Each NoSQL database has its own query language. 

NewSQL databases are fairly new and they combine features of both NoSQL and SQL, a new kind of Cloud database. They are "relational databases, supporting sharding, automatic replication, and distributed transaction processing" \cite{14}. A few examples of this are NuoDB, CockroachDB, VoltDB, and Clustrix, and most importantly Google Cloud Spanner.

Spanner is a globally, synchronized distributed database by Google. Spanner was made to be very scalable, and is designed to scale up to millions of machines \cite{15}. One advantage of Spanner is that it is available as part of the Google Cloud platform. The biggest advantage that spanner adds is the feature of high availability by allowing clients to replicate their data across continents. This is done by having global committing timestamps to transactions. This feature is accomplished by using an atomic clock. Google created the TrueTime API \cite{18} which uses atomic clocks and GPS receivers in each spanner data center to keep time consistent between its servers. One big disadvantage of this system however is in the latencies, a critical concern with distributed databases. 

Attested to this disadvantage, Bulanov \cite{16} performed an evaluation based on Yahoo! Cloud Serving Benchmark (YCSB) \cite{17}. In this study, three databases including NuoDB, CockroachDB, and Google Cloud Spanner were compared. For the experiment, they ran the YCSB Spanner tests on Google Cloud for Cloud Spanner, while benchmarks for CockroachDB and NuoDB resided on bare metal in their lab using 32GB, 4 core SuperMicros with 10G network.

In YCSB tests, there are a number of workloads named from A to E. In short, the workloads can be described as follows:\\

A - Heavy update (50\% read, 50\% update)

B - Mostly read (95\% read, 5\% update)

C - Read-only (100\% read)

D - Read the latest inserted (90\% read, 10\% insert)

E - Scan the latest inserted (90\% read, 10\% insert)

F - Read-modify-write (50\% read, 50\% update)\\

For each of the workloads, the application load (number of threads ran by YCSB application) was varied, and so was database capacity (number of nodes between 1 and 3). Multiple runs were executed, then the latencies numbers across all runs for each database were gathered. Fig. 1 and Fig. 2 show the results.

\begin{figure}
	\centering
	\includegraphics[height=6.2cm]{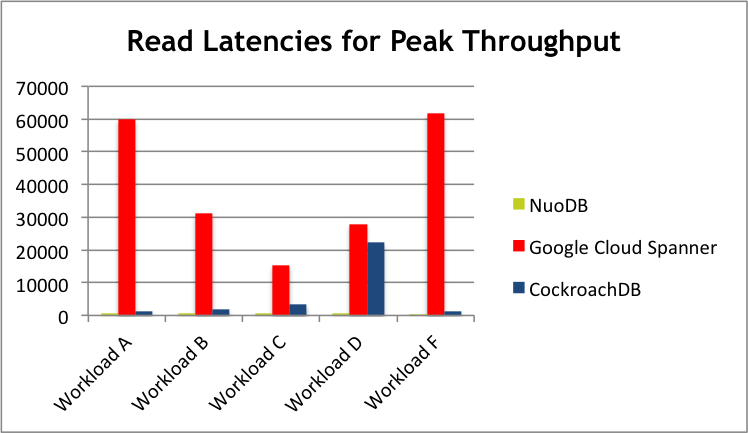}
	\caption{Latency experienced by the applications for READs}
	\label{fig1}
\end{figure}

The graph in Fig.1 measures average latency experienced by the application for READs during periods of peak throughput. As it can be seen from the figure, NuoDB's latency numbers were significantly lower than latency for Google Spanner and CockroachDB (minimal latency is ideal for the best user experience).

\begin{figure}
	\centering
	\includegraphics[height=6.2cm]{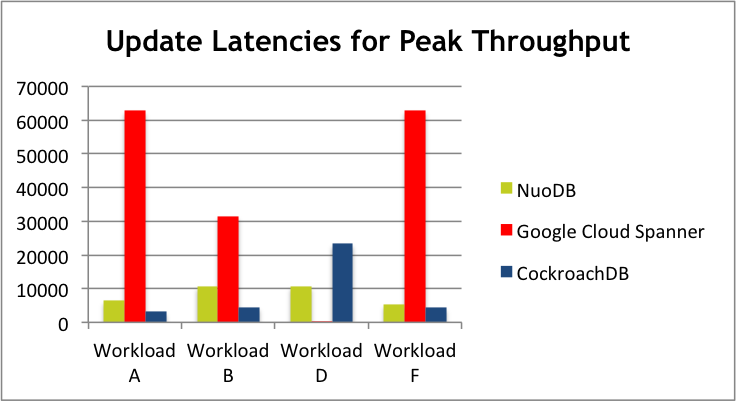}
	\caption{Latency experienced by the applications for UPDATEs and INSERTs}
	\label{fig2}
\end{figure}

The graph in Fig. 2 measures average latency experienced by the application for UPDATEs and INSERTs during periods of peak throughput. Note that Workload C was not represented here as it is a READ-only workload. As it can be seen from the figure, this was another area where NuoDB outperformed other databases.

\textit{As mentioned, having each microservice be its own component allows developers more freedom in choosing ways to store their data, satisfying a more purpose-fit technology stack practice. Additionally, microservices seem more technically in line with new data storage technologies, thus they could be used as effective components in terms of data storage supporting capability.}

\section{Discussion}
In addition to results presented in Section 3, we discuss briefly the main observations with respect to research questions, and outline a few directions and points for further research in this section.

\textit{RQ1. What are the main motivations to study microservices in CBSD?}

Microservices have become an effective means to design systems that can accommodate an expansive scale, an expansive domain, or a rapid release schedule, while keep a high degree of reusability. The demands of software systems today often expect one or many of these concerns to be met, making microservices a relevant tool for the design of component-based software systems. Particularly its scalability, which CBSD is severely lacking in it, is considered ideal when the system must enable support for a range of devices and platforms, including Web, Mobile, IoT, wearables, or simply when the users are not sure what kind of devices they will need to support in the future. There exists a myriad number of deployment and operation mechanisms centered around microservices. Innovation is near constant, as more are being added to the microservices' list daily (more efficient techniques and tools are on the rise)

\textit{RQ2. What are the limitations of microservices in this field?}

A microservice is a service that completes a business objective and it only handles that one specific function and uses messages as a way to communicate with each other. A microservice has to be independent leading to high cohesion and low coupling. The work in \cite{19} provides benchmark requirements. Even when the authors compared their benchmarks with software repositories, it showed that there is no consensus on all the requirements of what makes a microservice and its architecture, posing a serious adoption limitation for users and clients.

On a more specific level, the limiting factors that microservices provide are: The first being that they use a distributed system that increases modularity but also causes performance issues as remote calls are often slow and produce horrible latency. Another problem is its reliability to which remote calls can fail and need to be handled in a certain way. Consistency becomes a problem (as explained in the CAP theorem) when microservices need to update it may need multiple resources to do so.

Though these limitations might come at a price, but the amazing possibilities that microservices offer as components should be considered as factors related to decisions to adopt this architectural style, and how the likelihood of its utilization can be beneficial in a context-by-context basis.

\textit{RQ3. What should be the future research in this field?}

Implementation of a microservice-based system as components is still a difficult ordeal that requires a great deal of coordination, planning, and expertise. Since this field is still a new frontier, design patterns and best practices are still largely up for debate. Development of microservices could benefit greatly from the greater establishment and standardization of these concepts.

As another observation, we noticed is how few empirical studies have been conducted in microservices domain (and literally none for investigation of microservices as components) and how many cannot even be reproduced. Independent replication of interesting studies supported by valid meta-analysis is a good way to start. Software service engineering is not any area like physics or medicine where researchers vie to be first with some new idea, extending and improving a useful idea is valuable.

\section{Conclusions}
With the rise in popularity of microservices and a shift from large monolithic applications to small independent services, we believe that component-based software engineering will pick up on this. Much of the concepts of one are very interconnected with the other such as using a service that allows users access to a system. Service-oriented architecture (SOA) represents a good bridge between the two as a service that is being called on could be made by a collection of microservices. We approached this study with research question: can microservices be used as components in massively deployed and complex applications? We investigated the existing literature and reviewed a sample of the techniques and methods from selected researches, analyzing the core properties of microservices including scalability, consistency, availability, coupling and cohesion, and data storage technologies. As a general result, we conclude that using microservices as components is promising and would be a good mechanism for building applications that were used to be built with component-based approaches.

We hope this work can be useful in the process of building a more community spirit to study CBSD from a new perspective, since microservices field is evolving and potentially has much to offer to revive component-based approaches.


\begin{thebibliography}{4}


\bibitem{1} P. D. Francesco, I. Malavolta and P. Lago, Research on Architecting Microservices: Trends, Focus, and Potential for Industrial Adoption, Proceedings of the 2017 IEEE International Conference on Software Architecture (ICSA17), pp. 21-30 (2017).

\bibitem{2}	E. S. de Almeida, A. Alvaro, D. Lucredio, A. F. do Prado and L. C. Trevelin, Distributed component-based software development: an incremental approach, Proceedings of the 28th Annual International Computer Software and Applications Conference (COMPSAC 2004), pp. 4-9 (2004)

\bibitem{3} D. McIlroy, Mass-produced software components, NATO Conference, pp. 88-98 (1968)

\bibitem{4} T. Vale, I. Crnkovic, E. S. de Almeida, P. A. da Mota Silveira Neto, Y. C. Cavalcanti, S. R. de Lemos Meira, Twenty-eight years of component-based software engineering, Journal of Systems and Software, Vol. 111, pp. 128-148 (2016)


\bibitem{5} N. Dragoni et al. (2017) Microservices: Yesterday, Today, and Tomorrow. In: Mazzara M., Meyer B. (eds) Present and Ulterior Software Engineering. Springer.


\bibitem{6} D. Shadija, M. Rezai and R. Hill, Towards an understanding of microservices, Proceedings of 23rd International Conference on Automation and Computing (ICAC), Huddersfield, pp. 1-6 (2017)


\bibitem{7} N. Alshuqayran, N. Ali, R. Evans (2016) A Systematic Mapping Study in Microservice Architecture, Proceedings of IEEE 9th International Conference on Service-Oriented Computing and Applications, Macau, China, November 4-6, 2016.

\bibitem{8} A. R. Sampaio et al., Supporting Microservice Evolution, Proceedings of 2017 IEEE International Conference on Software Maintenance and Evolution (ICSME), Shanghai, 2017, pp. 539-543.

\bibitem{9} T. Cerny, M. J. Donahoo, and M. Trnka, Contextual understanding of microservice architecture: current and future directions, ACM SIGAPP Applied Computing Review,  vol. 17, pp. 29-45, 2018.


\bibitem{10} M. Schwarz, "Uber Engineering's Micro Deploy: Deploying Daily with Confidence," Uber Engineering, June, 23, 2016.

\bibitem{11} S. Gilbert and N. Lynch, "Perspectives on the CAP Theorem," in Computer, vol. 45, no. 2, pp. 30-36, 2012.


\bibitem{12} R. Heffner, "Divide and Conquer in Software Architecture," Computer Weekly, pp. 15-17, 2016.

\bibitem{13} M. Fowler, "Microservice Trade-Offs," https://martinfowler.com/articles/microservice-trade-offs.html, 2015.

\bibitem{14} A. Corbellini, C. Mateos, A. Zunino, D. Godoy, and S. Schiaffino, "Persisting big-data: The NoSQL landscape," Information Systems, Article vol. 63, pp. 1-23,2017.

\bibitem{15} J. C. Corbett et al., "Spanner: Google's Globally Distributed Database," ACM Transactions on Computer Systems, vol. 31, no. 3, pp. 1-22, 2013.

\bibitem{16} B. Bulanov, "Benchmarking Google Cloud Spanner, CockroachDB, and NuoDB," September, 21 2017.
\bibitem{17} Brian F. Cooper, Yahoo! Cloud Serving Benchmark (YCSB), https://github.com/brianfrankcooper/YCSB/wiki, Feb 27, 2018.

\bibitem{18} E. Brewer, "Spanner, TrueTime and The CAP Theorem," Research at Google, February,14 2017.

\bibitem{19} C. M. Aderaldo, N. C. Mendonça, C. Pahl and P. Jamshidi, "Benchmark Requirements for Microservices Architecture Research," 2017 IEEE/ACM 1st International Workshop on Establishing the Community-Wide Infrastructure for Architecture-Based Software Engineering (ECASE), Buenos Aires, 2017, pp. 8-13.



\end{thebibliography}
\end{document}